\documentclass[11pt,letterpaper]{article}
\usepackage{stmaryrd}
\usepackage[dvips]{graphicx}
\usepackage{amssymb,amsmath,amsfonts,palatino,amsthm}
\usepackage{amssymb,mathtools}
\usepackage[sans]{dsfont}
\usepackage{booktabs}
\usepackage{psfrag}

\newcommand\rmax{r_\mathrm{max}}
\newcommand\tmax{t_\mathrm{max}}

\newcommand{\be}{\begin{displaymath}}
\newcommand{\ee}{\end{displaymath}}
\newcommand{\bne}{\begin{equation}}
\newcommand{\ene}{\end{equation}}
\newcommand{\eq}[1]{Eq.\ (\ref{#1})}
\newcommand{\cqg}{\emph{Class.\ Quant.\ Grav.}}
\newcommand{\eprint}[1]{$\langle$e-print arXiv: #1$\rangle$}

\newcommand{\cmt}[2]{\mbox{\begin{minipage}[t]{#1cm} #2 \end{minipage}}}

\begin{document}

\title{A Causal Set Black Hole}
\author{Song He\thanks{Email address:
hesong@pku.edu.cn}\\ \\School of Physics, Peking University,
Beijing, 100871, China, \\ \\ \\David Rideout\thanks{Email address:
drideout@perimeterinstitute.ca}
\\ \\
Perimeter Institute for Theoretical Physics,\\
31 Caroline St.\ N, Waterloo, Ontario N2L 2Y5, Canada}

\maketitle
\vfill
\begin{abstract}
We explicitly compute the causal structure of the Schwarzschild black hole
spacetime, by providing an algorithm to decide if any pair of events is
causally related.  The primary motivation for this study comes from discrete
quantum gravity, in particular the causal set approach, in which the
fundamental variables can be thought of as the causal ordering of randomly
selected events in spacetime.  This work opens the way to simulating
non-conformally flat spacetimes within the causal set approach, which may
allow one to study important questions such as black hole entropy and Hawking
radiation on a full four dimensional causal set black hole.

\end{abstract}
\vfill
\tableofcontents

\section{Introduction}
\label{intro}

Much is understood about the causal structure of the Schwarzschild black hole
spacetime, e.g.\ in the sense that the collection of all null geodesics has
been completely classified \cite{Chand}.  However, what is missing is a
complete specification of the causal \emph{relations}, namely the set of
ordered pairs of events in Schwarzschild which are connected by a causal curve.
It is the purpose of this paper to provide exactly such a prescription:
\emph{Given an arbitrary ordered pair of events in Schwarzschild spacetime,
  does there exist a future directed causal curve from the first to the
  second?}

Unfortunately, the differential equations describing the null geodesics are
solvable in closed form only for limited values of their parameters
\cite{Chand}.  (In fact only a set of measure zero are expressible in closed
form.)  We therefore describe an algorithm
which will allow one to compute an answer to the above question, to any
(finite) precision.
For more implicit discussion on how the space of null geodesics encodes the 
causal structure of Schwarzschild, and general spacetimes,
please refer to \cite{low,low_schwarzschild} and references therein.\\

The primary motivation for this work comes from 
discrete quantum gravity, in
particular the causal set approach, for which the fundamental variables can be
regarded as the causal ordering of randomly selected events in spacetime
\cite{blms, myrheim, random_discrete}.
More specifically the causal set is a discrete set of `atoms of spacetime',
which possesses a partial order relation which corresponds to causal ordering
in spacetime.
Because of this straightforward interpretation of the fundamental variables,
it is relatively easy to extract phenomenological predictions from the theory
on the effects of fundamental spacetime discreteness.  Perhaps the most famous of
these is the prediction of a small but non-zero fluctuating cosmological
constant, whose current order of magnitude matches its observed value
\cite{discrete_lambda}.

Mathematically a causal set is a set $C$ endowed with an order
relation $\prec$ with is irreflexive ($x \nprec x$), transitive ($x \prec y$
and $y \prec z \implies x \prec z$), and locally finite ($|\{y | x \prec y
\prec z\}|$ is finite for all $x,z \in C$)\footnote{The vertical brackets $|
  \cdot |$ stand for cardinality.}.
The connection between a microscopic discrete causal set and a
macroscopic approximating continuum spacetime arises via the notion of a
`sprinkling', which is a simple algorithm for generating a causal set from a spacetime.
Given a spacetime $M$ with finite spacetime volume $V$ (such as a bounded region of
an infinite spacetime), select at random $N$ events in $M$, with respect to
the volume measure.  By this we mean that in any region of $M$ of volume $v$
one expects to find $n$ sprinkled events, where $n$ is sampled from a
Poisson distribution of mean $v$.  Thus the probability of finding $n$
sprinkled events in a region of volume $v$ is
\bne
Pr(n,v) = \frac{v^n e^{-v}}{n!} \;.
\label{poisson}
\ene
(Therefore $N$ is sampled from the distribution \eq{poisson} with $v=V$.)
Each of these sprinkled events then corresponds to an element of the causal set.
After sprinkling the $N$ events, one defines the partial ordering by stating
that two elements are related iff the events are causally related in the
spacetime.
The microscopic--continuum correspondence then arises by the statement that if
a causal set is likely to have arisen from a sprinkling into a given
spacetime, then one regards that spacetime as a good macroscopic
approximation of the underlying causal set.

Thus far only conformally flat spacetimes, viz.\ those whose metric is given by a scalar conformal times the Minkowski metric, have been used in detailed
calculations involving sprinklings of
causal sets, because the second step of deducing the causal relations is
particularly easy in that case.  Here we open the possibility for sprinkling
into a non-conformally flat spacetime: the Schwarzschild black hole.  This
allows one to address the general question of whether the methods of deducing
properties of continuum spacetime from the causal set carry over to curved
spacetime, and in particular spacetime with non-flat conformal structure.
Some constructs which one may like to test are dimension estimators \cite{myrheim,dimension}, the
recovery of lengths of timelike
\cite{random_discrete,timelike_dist} and spacelike \cite{spatial_dist} geodesics,
and extraction of macroscopic spatial topology \cite{homology}.

The black hole is important for a number of other reasons, beyond merely
being an example of a non-conformally flat spacetime.  In the same way in
which one can get a first approximation of the entropy of a gas merely by
counting molecules, there is now substantial 
evidence that one can do an
analogous counting of `horizon molecules', such as causal links of a causal set
crossing the horizon, to compute the entanglement entropy of a black hole, and thus get a
handle on its physical origin \cite{bhentropy}.
In addition one may test a recently proposed entropy bound
\cite{entropy_bound} in this spherically symmetric,
yet conformally curved context.
Finally, this work can shed light on the longstanding problem of the role
trans-Planckian modes in Hawking radiation, by allowing the study of a full
four dimensional causal set black hole.\\

The technique we present here, of deducing the causal relation from a
classification of the null geodesics of a spacetime, can be generalized to
other spacetimes as well.  The most obvious example is the Reissner-Nordstrom
black hole.  It remains spherically symmetric, which considerably simplifies
the analysis, and its null geodesics are completely classified \cite{Chand}.
Presumably one could also analyze the Kerr black hole in this way, though the
broken spherical symmetry will be considerably more complicated.  Note also
that one must avoid any regions which contain closed timelike curves (or
consider a more general ordered substructure),
since the causal set description breaks down there.\\

In addition to describing an algorithm 
to decide if any two events in
Schwarzschild are connected by a causal curve, we implement this
prescription as a `thorn' (module) in the Cactus high performance computing framework
\cite{cactus}.  An advantage of doing so is that others can easily make use
of the code, without having to write their own implementation of the
algorithm, nor having to understand the details of the representation of the
causal set on the computer.\\

This paper is organized as follows.  In section \ref{null_geodesics} we
briefly describe the equations governing the null geodesics we will employ,
and describe how to use them to determine if two events are causally related or not.
In section \ref{results} we present some of the causal
relations as computed by our algorithm, and pictures which
illustrate the causal sets which arise by sprinkling into a region of
Schwarzschild spacetime.  Section \ref{conclusion} contains some concluding
remarks.  In Appendix \ref{proof} we give a proof that the null geodesics we
integrate to determine the causal relations are those that arrive earliest.
In Appendix \ref{details}
we describe the details of our implementation of
this method,
including how to sprinkle into the Schwarzschild geometry with
uniform density.

\section{Null geodesics and the causal structure of a Schwarzschild black hole}
\label{null_geodesics}

\subsection{Preliminaries}
\label{lemma_section}
In this paper, we want to find a general recipe to determine
unambiguously whether two events in four dimensional Schwarzschild
spacetime are causally related to each other. This problem has a
simple answer in Minkowski spacetime, since it is straightforward to
show that if
$-(t_2-t_1)^2+(x_1-x_2)^2+(y_1-y_2)^2+(z_1-z_2)^2\leq0$, two events
$E_1=(t_1,x_1,y_1,z_1)$ and $E_2=(t_2,x_2,y_2,z_2)$ are causally related, otherwise there is no causal
relation between them. Things become more complicated in curved
spacetime, where in principle we need to integrate the infinitesimal
invariant distance $ds$ along every possible path from one event to
the other to see if there is a null or timelike curve (causal curve)
connecting them.

\begin{figure}[htbp]
\center
\psfrag{g}{\Large $\gamma$}
\psfrag{E1}{$E_1$}
\psfrag{E2}{$E_2$}
\includegraphics[height=10cm]{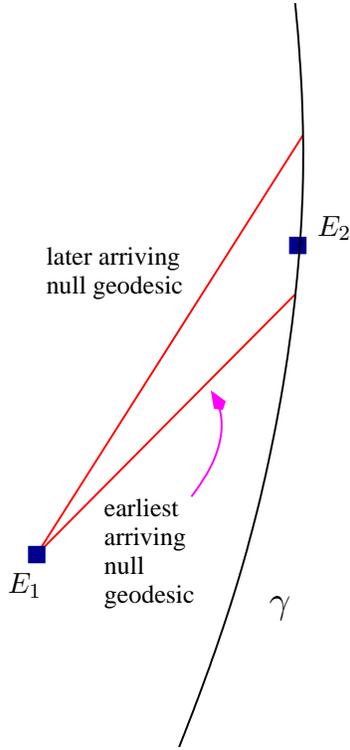}
\caption{Illustration of the timelike worldline $\gamma$, on which all
  spatial coordinates (including $r$) are constant, and two null geodesics
  from $E_1$ to $\gamma$.}
\label{worldline}
\end{figure}
Nevertheless, it is possible to solve
this
problem, given an understanding of the collection of null geodesics in the
spacetime.  In the Schwarzchild metric, it is always
possible to choose a time parameter that only increases towards
future, i.e.\ $\partial_t^a$
is everywhere a future pointing timelike vector.\footnote{We will choose Eddington-Finkelstein coordinates,
which have a such a time parameter, below.}  Then given two events
$E_1$ and $E_2$ with $t_1<t_2$, the only possible causal relation
between them is that $E_1\prec E_2$, which means that there is a
future-directed causal curve from $E_1$ to $E_2$. Imagine a bunch of
light rays (null geodesics) $B$, emanating from $E_1$, and $E_2$ represents a
particular moment $t_2$ in the world line $\gamma$ of a stationary
observer. If the world line of any null geodesics in $B$, which is a null
geodesic, meets $\gamma$ at $t\leq t_2$, then we can conclude that
they are causally-related. On the other hand, if any null geodesic
emanating from $E_1$ can only reach $\gamma$ at $t>t_2$, then $E_1$
and $E_2$ must be causally unrelated to each other.  See
Fig.\ \ref{worldline} for an illustration.  That this is true can be seen by
proposition 2.20 of \cite{Penrose}, which states if $A \in J^-(B)$ but $A
\not\in I^-(B)$, then there must exist a future-directed null geodesic from $A$ to
$B$.
This tells us
that the earliest future-directed causal curve must be
a null geodesic, but since any null geodesic has failed to reach
$\gamma$ early enough for a causal relation between the two events,
then the conclusion is that there is no causal relation between
them.

We will prove in Appendix \ref{proof} that our procedure always considers the
fastest\footnote{By `fastest' we mean the geodesic with the earliest arrival
  time, as given by the (EF) time coordinate.}
geodesic from $E_1$ to $\gamma$, so the arrival time $t$ at $\gamma$ of that
geodesic will be sufficient to determine if $E_1$ and $E_2$ are causally
related.
If we find that even this null geodesic meets $\gamma$
later than $t_2$, we can say for sure that there is no way for these two
events to be connected by any future directed causal curve.

After this brief introduction, 
we discuss in the next
subsection the particular simple case where $E_1$ and $E_2$ are only
radially separated, with no angular separation, so all we need to consider are radial
geodesics. For more generic pairs of events, we must consider
the full three dimensional case,\footnote{Given two events in Schwarzschild
  spacetime, is is always possible to rotate the coordinates so that they
  both lie in the equatorial plane, as explained below.}
which is too complicated for a complete
analytic solution. However, before incorporating a numerical treatment
for the generic case, in subsection \ref{sufficient condition} we
find two simple sufficient conditions for 
two events to be
causally unrelated to each other.  One uses a bound 
given by radial null geodesics, and the other by purely angular
components. Furthermore, there is also a sufficient condition for
two events to have causal relation, which is the existence of a
composed null curve connecting $E_1$ and an event in $\gamma$ no
later than $E_2$. These sufficient conditions are enough to
determine the causal relations for a large portion of pairs of
events in Schwarzschild spacetime, and provide a very efficient
preconditioning, since we only need to do numerical calculations for
those which fail all these conditions. Furthermore, we shall give
the recipe for generic pairs of events in subsection \ref{generic
pairs}, with the help of numerical
calculations. As mentioned above, the proof of the 
lemma which ensures that the
null geodesic found by our recipe is the fastest one connecting
$E_1$ and $\gamma$ is given in the Appendix.\\

We start with the Schwarzschild metric in the familiar form
\be 
ds^2=-(1-\frac{2M}{r})dt_s^2+(1-\frac{2M}{r})^{-1}dr^2+r^2(d\theta^2+\sin^2\theta
d\phi^2) \;,
\ee 
where $M$ is the mass of the black hole. The metric is well known to
possess a coordinate singularity at $r=2M$, where the event horizon lies.  It can be
written in Eddington-Finkelstein (EF) coordinates \cite{Possion},
\begin{equation}\label{EF}
ds^2=-dt^2+dr^2+r^2(d\theta^2+\sin^2\theta
d\phi^2)+\frac{2M}{r}(dt+dr)^2 \;,
\end{equation}
with the following transformation of time parameter,
\begin{equation}\label{Transform}
t=t_s+2M\ln(\frac{r}{2M}-1) \;.
\end{equation}

Given two events in the EF coordinates, $E_1=(t_1,r_1,\theta_1,\phi_1)$ and
$E_2=(t_2,r_2,\theta_2,\phi_2)$ with $t_1\leq t_2$, the only
possible causal relation is $E_1 \prec E_2$. Besides, it is
obvious that one can always choose suitable angular coordinates 
$\vartheta$ and $\varphi$ 
for which $\vartheta_1=\vartheta_2=\pi/2$,
$\varphi_1=0$, and
$\varphi_2=\arccos(\cos\theta_1\cos\theta_2+\sin\theta_1\sin\theta_2\cos(\phi_1-\phi_2))\in[0,\pi]$.
Therefore, it is sufficient to consider a pair of events
$E_1=(t_1,r_1,\pi/2,0)$ and $E_2=(t_2,r_2,\pi/2,\varphi_2\in[0,\pi])$
with $t_1\leq t_2$.
The stationary worldline $\gamma$ containing $E_2$ is given by $(t, r=r_2,
\vartheta=\frac{\pi}{2}, \varphi=\varphi_2)$, i.e.\ all spatial coordinates are held
fixed.
We define the angle through which our null geodesic travels, $\Delta \varphi$,
by
\be
\Delta \varphi = \int_{\eta} d\varphi \;,
\ee
where $\eta$ is a null geodesic from $E_1$ to $\gamma$, and
$\varphi$ is the azimuthal coordinate in the rotated coordinate system.
In subsection \ref{generic pairs} we shall find that a generic null
geodesic from $E_1$ to $\gamma$ can always be formulated as one with
$\vartheta=\pi/2$ as well, thus the whole problem, including the events
and geodesics in between, can be projected to $2+1$ dimensions.

An important question when considering the generic pair of events will be if
the null geodesic we choose is the fastest one, i.e.\ that it arrives at
$\gamma$ before all other null geodesics from $E_1$.  In order for the null
geodesic $\eta$ to arrive at $\gamma$, it must have $\Delta \varphi = 2k\pi +
\varphi_2$, with $k \in \mathbb{Z}$.
In Appendix \ref{proof} we
prove that \emph{a geodesic $\eta$ with $|\Delta \varphi| = \varphi_2$ is
  a fastest 
  future directed null geodesic from $E_1$ to $\gamma$.}\footnote{The absolute value on $|\Delta \varphi|$ is merely to account for
  the case of $\varphi_2 = \pi$, in which there are two fastest null
  geodesics, i.e.\ that both have the same arrival time.  However, note that
  this case will never arise in the sprinkling described in Section
  \ref{intro}, as the set of such sprinklings is of measure zero in the space
  of all sprinklings.}
Thus we only need to consider
those null geodesics from $E_1$ to $\gamma$ which travel
for an angle no more than $\pi$ in the $\varphi$ direction.

\subsection{Radially separated pairs and radial null geodesics}\label{radial}

\begin{figure}[htbp]
\center
\psfrag{t}{$t_s$}
\psfrag{r}{$r$}
\includegraphics[height=6cm]{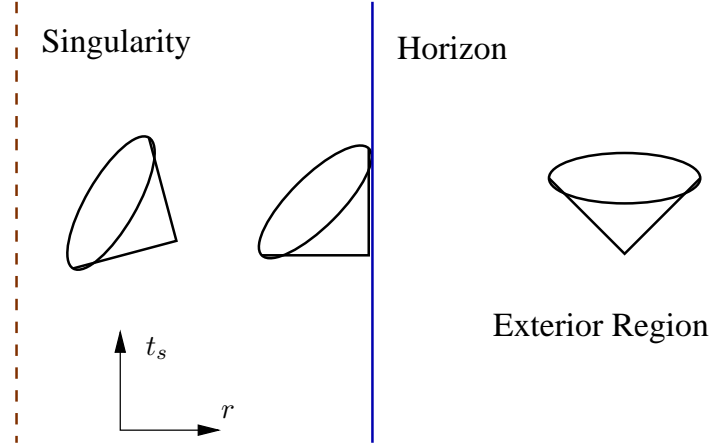}
\caption{Schematic illustration of lightcones in a Schwarzschild black hole.
  The event horizon is indicated by the blue line, and the interior region is
  to the left.  Note the two ingoing radial directions for future directed
  lightcones in the interior region.  (The vertical direction is to be
  regarded here as Schwarzschild time, which can go backward for timelike
  observers inside the horizon.)}
\label{lightcones}
\end{figure}

Now let us consider the simplest case with $\varphi_2=0$.
In this case we only need to consider radial null geodesics, and it
is straightforward that in the EF coordinates, by setting
$d\vartheta=d\varphi=0$ and $ds^2=0$, radial null geodesics take a simple
form,
\begin{equation}\label{Radialnullgeoingo}
dt+dr=0 \;,
\end{equation}
\begin{equation}\label{Radialnullgeo}
\left( \frac{2M}{r}-1 \right) dt + \left( \frac{2M}{r}+1 \right) dr = 0 \;,
\end{equation}
where the first one describes ingoing null geodesics, while the second
one turns out to be outgoing for null geodesics outside the event
horizon, and ingoing for those inside the event horizon.  The two radial
directions for
null geodesics, both outside and inside the horizon, are illustrated in Fig.\ \ref{lightcones}.
Therefore,
given two events without angular separation, $E_1=(t_1,r_1,\pi/2,0)$
and $E_2=(t_2,r_2,\pi/2,0)$ with $t_1\leq t_2$, we can determine if
they are causally related by considering the time for null geodesics
from $E_1$ to reach a point in $\gamma$, the world line of a
stationary observer at $(r_2,\pi/2,0)$.

First assume $r_1\geq r_2$.  We consider
only ingoing null geodesics
from $r_1$ to $r_2$. If both events are outside the horizon,
$r_1\geq r_2>2M$, there are only ingoing null geodesics given by
Eq.\ (\ref{Radialnullgeoingo}), and it is straightforward to see that
these two events are causally related to each other if and only if
$t_2\geq t_1+r_1-r_2$, where $t_1+r_1-r_2$ is simply the time when a
null geodesic from $E_1$ hits $\gamma$. If $r_1\geq 2M\geq r_2$, we still
only have Eq.\ (\ref{Radialnullgeoingo}) for ingoing rays outside the
horizon, and to ensure continuity of the tangent 
of the
geodesic when it crosses the horizon, we can only use
Eq.\ (\ref{Radialnullgeoingo}) inside the horizon too, and the
conclusion is not altered. However, if both events are inside the
horizon, $2M\geq r_1\geq r_2$, we have two choices of ingoing rays given
by Eq.\ (\ref{Radialnullgeoingo}) and Eq.\ (\ref{Radialnullgeo}), and a
simple calculation shows that it takes a shorter time for a null geodesic
from $E_1$ to hit $\gamma$ for the first case, $dt=-dr=|dr|$, than
for the second case
$dt=-\frac{2M/r+1}{2M/r-1}dr=(1+\frac{2r}{2M-r})|dr|>|dr|$ for
$r<2M$. Therefore, for both events inside the horizon, the necessary
and sufficient condition for their causal relation is still $t_2\geq
t_1+r_1-r_2$.

On the other hand, if $r_2\geq r_1$, we shall instead consider outgoing
rays, given by Eq.\ (\ref{Radialnullgeo}) outside the horizon. There
is a special case, $r_1<2M$, for which no matter $r_2\leq 2M$ or
$r_2>2M$, they can not be causally related to each other for any
$t_1$ and $t_2$, since there are no outgoing null geodesics either
propagating inside the horizon or coming out of it.  If they both lie
outside the horizon, $r_2\geq r_1>2M$ , then the sufficient and
necessary condition for a causal relation between them is $t_2\geq
t_1+r_2-r_1+4M\ln(\frac{r_2-2M}{r_1-2M})$, where
$t_1+r_2-r_1+4M\ln(\frac{r_2-2M}{r_1-2M})=t_1+\int_{r_1}^{r_2}\frac{r+2M}{r-2M}dr$
is the time for a null geodesic given by Eq.\ (\ref{Radialnullgeo}) to
travel from $E_1$ to the point it meets $\gamma$.

To summarize, we have the following recipe to determine if two
events are causally related to each other.  If $r_1\geq r_2$, the
(necessary and sufficient) condition for a causal relation is always
\begin{displaymath} t_2\geq t_1+r_1-r_2 \;, \end{displaymath} while if $r_2\geq r_1>2M$, they are
related if and only if \begin{displaymath} t_2\geq
t_1+r_2-r_1+4M\ln(\frac{r_2-2M}{r_1-2M}) \;. \end{displaymath}
Finally if $r_2>r_1$ and $r_1<2M$, they can not be causally related
to each other.

\subsection{Sufficient conditions for causally related and unrelated
  pairs}\label{sufficient condition}

Since in practice we are interested in computing the causal relations between
many pairs of events, it is useful to derive bounds that allow one to decide
whether the events are related quickly, based on simple criteria, without having to perform any numerical
integrations.  In this section we derive such bounds, which are sufficient to
determine that the two events are unrelated (section \ref{spacelike-sec}) or
related (section \ref{timelike_bnd}).

\subsubsection{Spacelike bounds}
\label{spacelike-sec}
For two events with non-zero angular distance, i.e.\
$\varphi_2\in(0,\pi]$, we have to consider non-radial, three
dimensional null geodesics in general \cite{Chand}. Nevertheless, we
shall find two independent sufficient conditions for the events to be
causally unrelated below, by using lower bounds on the time duration
of any future-directed causal curve from $E_1$ to $\gamma$, given by
either radial null geodesics with the angular part discarded, or by a purely
angular component with the radial part discarded.

Since anywhere on a future-directed causal curve we have $ds^2\leq0$
and $dt>0$, with \eq{EF} we obtain an inequality by
discarding the (positive) angular part,
\be 
(dt+dr)\left[\left(\frac{2M}{r}-1\right)dt+\left(\frac{2M}{r}+1\right)dr\right]\leq 0 \;.
\ee 
If $r>2M$, we obtain for an ingoing causal curve $dr<0$, $dt\geq-dr$, and
for an outgoing curve $dr>0$, $dt\geq\frac{r+2M}{r-2M}dr$. If $r<2M$, we
obtain for an ingoing curve $dr<0$, $\frac{r+2M}{r-2M}dr\geq dt\geq-dr$, and no solution
for an outgoing curve $dr>0$.

These inequalities can be integrated along the whole curve from
$E_1$ to $\gamma$ to give a lower bound on the time duration, which
turns out to be exactly the time duration of radial null geodesics
from $r_1$ to $r_2$. If $r_1\geq r_2$, the sufficient condition for
two events to be causally unrelated is
\begin{equation}
t_2 - t_1 < r_1-r_2 \;,
\label{rad_bnd_ingoing}
\end{equation}
while if $r_2\geq r_1>2M$, they are unrelated if
\begin{equation}
t_2 - t_1 < r_2-r_1+4M\ln(\frac{r_2-2M}{r_1-2M}) \;.
\label{rad_bnd_outgoing}
\end{equation}
Finally if $r_2>r_1$ and $r_1<2M$, they can not be causally related
to each other.

On the other hand, we can also try to discard the radial part to obtain
a bound by using a purely angular component, but this is more subtle
and we need to be careful. Since $dr^2+\frac{2M}{r}(dt+dr)^2\geq0$,
we can discard this part to obtain an inequality anywhere on a
causal curve, as $-dt^2+r^2d\varphi^2\leq0$, which gives $dt\geq
r|d\varphi|$. However, for outgoing causal curves outside the
horizon, we have $(1+\frac{2M}{r})dr^2+\frac{4M}{r}dtdr>0$ and a
stronger bound can be obtained as
$(\frac{2M}{r}-1)dt^2+r^2d\varphi^2\leq0$ which gives
$dt\geq\frac{r|d\varphi|}{\sqrt{1-2M/r}}$, where $r>2M$. The
equality holds for null geodesics with constant $r$, which
satisfy
\begin{equation}\label{Angularnullgeo}
\sqrt{1-\frac{2M}{r}}dt\pm rd\varphi=0 \;,
\end{equation}
where $r=\mathrm{const}>2M$.
This can be rephrased physically by stating that it is impossible
for a particle inside the horizon to move around an orbit with
constant $r<2M$, because nothing can stop it from falling into the
singularity $r=0$. We shall make direct use of
Eq.\ (\ref{Angularnullgeo}) later for the timelike bound.

For any ingoing curve from $E_1$ to $\gamma$, we have $dt\geq
r|d\varphi|\geq r_2|d\varphi|$ along it, and we obtain a sufficient
condition for two events with $r_1\geq r_2$ to be causally
unrelated, \be t_2<t_1+r_2\varphi_2 \;. \ee

For any outgoing curves from $E_1$ to $\gamma$, we have
$dt\geq\frac{r|d\varphi|}{\sqrt{1-2M/r}}$ along it, and we need to
find out the minimum of $f(r)=\frac{r}{\sqrt{1-2M/r}}$ in the range
$2M<r_1\leq r\leq r_2$.  It is straightforward to obtain
\be 
f'(r)=\frac{1-3M/r}{(1-2M/r)^{3/2}} \;,
\ee 
from which the location of the minimum can be determined to be
$r_0=r_1$ for $3M\leq r_1\leq r_2$, $r_0=3M$ for $r_1<3M< r_2$, and
$r_0=r_2$ for $r_1\leq r_2\leq 3M$.

Therefore we have $dt\geq\frac{r_0|d\varphi|}{\sqrt{1-2M/r_0}}$
along the curve and we obtain a sufficient condition for two events
with $2M<r_1\leq r_2$ to be causally unrelated,
\begin{equation}
t_2 - t_1 < f(r_0)\varphi_2 \;.
\label{ang_bnd}
\end{equation}

\subsubsection{Timelike bound}
\label{timelike_bnd}

Furthermore, it turns out that for many pairs, as long as at
least one of them is outside the horizon, we can use radial null
geodesics and null geodesics with constant $r$ to find a composed
null curve connecting $E_1$ and an event in $\gamma$.  If this event is no later than
$E_2$, then this will be a sufficient condition for their causal
relation.

Given Eq.\ (\ref{Radialnullgeoingo}), Eq.\ (\ref{Radialnullgeo}) and
Eq.\ (\ref{Angularnullgeo}), we can construct a null curve from $E_1$
to an event in $\gamma$ whenever $r_1>2M$, which is composed of
a sequence of null geodesics with constant $\varphi$ and constant $r$.
To optimize this sufficient condition for a causal relation, we need
to minimize the time duration of the segment of the null geodesic with
constant $r$, which 
occurs at $r=r_0$ where
$r_0=\mathrm{min}(r_1,r_2)$ for $r_1,r_2\geq 3M$, $r_0=3M$ for
$r_1>3M>r_2$ or $r_2>3M>r_1$, and $r_0=\mathrm{max}(r_1,r_2)$ for
$r_1,r_2\leq 3M$.

If $r_1\geq r_0\geq r_2$, then we can
compose the null curve by the following three segments:

\begin{enumerate}
\item an ingoing radial segment from $r=r_1$
to $r=r_0$ with $\varphi=0$;

\item a segment from $\varphi=0$ to
$\varphi=\varphi_2$ with $r=r_0$;

\item an ingoing radial segment from
$r=r_0$ to $r=r_2$ with $\varphi=\varphi_2$.
\end{enumerate}
The time for this null curve to reach $\gamma$ is easy to compute by Eq.\ (\ref{Radialnullgeoingo})
and Eq.\ (\ref{Angularnullgeo}),
\begin{equation}
t=t_1+r_1-r_2+f(r_0)\varphi_2 \;.
\end{equation}
Here we do not care
about whether $r_2$ is inside or outside the horizon, since regardless of
whether
the segment 3 lies completely outside the horizon or crosses 
the horizon, we always use Eq.\ (\ref{Radialnullgeoingo}) for
continuity. Similarly, if $r_2\geq r_0\geq r_1>2M$, then we can
construct the null curve by replacing segments 1 and 3 above by outgoing
radial segments, and the time can be computed by
Eq.\ (\ref{Radialnullgeo}) and Eq.\ (\ref{Angularnullgeo}) to be
\be 
t=t_1+r_2-r_1+4M\ln(\frac{r_2-2M}{r_1-2M})+f(r_0)\varphi_2 \;.
\ee 

Therefore, we have the following sufficient condition for two
events, of which at least one is outside the horizon, to be causally related.
If $r_1\geq r_2$ and $r_1>2M$, then they are causally related if
\be 
t_2\geq t_1+r_1-r_2+f(r_0)\varphi_2 \;;
\ee 
if $r_2\geq r_1> 2M$, they are causally related if
\be 
t_2\geq
t_1+r_2-r_1+4M\ln(\frac{r_2-2M}{r_1-2M})+f(r_0)\varphi_2 \;.
\ee 

If a pair of events fails both of these sufficient
conditions, then we have to consider a generic 
form of null geodesics to determine if they are causally related.

\subsection{Generic pairs of events and null geodesics}\label{generic pairs}

The most generic null geodesics in Schwarzschild spacetime have the
following form \cite{Chand},
\be 
p_t\frac{dt}{d\tau}-p_r\frac{dr}{d\tau}-p_\theta\frac{d\theta}{d\tau}-p_\phi\frac{d\phi}{d\tau}=0
\;,
\ee 
where $\tau$ is an affine parameter, and $p_t$, $p_\phi$ are
constants,
\begin{eqnarray}
p_t=(1-\frac{2M}{r})\frac{dt_s}{d\tau}&=&E \;, \nonumber\\
p_\phi=r^2\sin^2\theta\frac{d\phi}{d\tau}&=&L \;, \nonumber
\end{eqnarray}
and $p_\phi$ satisfies
\be 
\frac{d(r^2\frac{d\theta}{d\tau})}{d\tau}=r^2\sin\theta\cos\theta(\frac{d\phi}{d\tau})^2
\;.
\ee 
If we choose $\vartheta=\pi/2$ at the moment when
$\frac{d\vartheta}{d\tau}=0$, we get also $\frac{d^2\vartheta}{d\tau^2}=0$ at
this moment, which implies $\vartheta=\pi/2$ all along the geodesic.
Therefore a general null geodesic can be described in the plane
$\vartheta=\pi/2$, which also simplifies its equations to
\begin{equation}\label{Nullgeo}
(\frac{dr}{d\tau})^2+\frac{L^2}{r^2}(1-\frac{2M}{r})=E^2 \;,
\end{equation}
where $E$ and $L$ denote the constant energy and angular momentum of
the massless particle,
\begin{eqnarray}\label{Conserve}
(1-\frac{2M}{r})\frac{dt_s}{d\tau}&=&E \;,\nonumber\\
\frac{d\varphi}{d\tau}&=&\frac{L}{r^2} \;.
\end{eqnarray}
The energy is expressed in terms of the Schwarzschild time
parameter $t_s$.

The full set of Eq.\ (\ref{Nullgeo}) and Eq.\ (\ref{Conserve}),
combined with initial values of $t$, $r$ and $\varphi$, as well as $E$
and $L$, can uniquely determine a null geodesic in Schwarzschild
spacetime. However, since we only want to obtain relations between $t$, $r$
and $\varphi$ without the affine parameter $\tau$, it is convenient to
consider $r$ as a function of $\varphi$ and use a new variable $u=1/r$.
We then obtain from Eq.\ (\ref{Nullgeo}),
\be 
(\frac{du}{d\varphi})^2=2Mu^3-u^2+c^2 \;,
\ee 
or equivalently,
\begin{equation}\label{Uphi}
\frac{d\varphi}{du}=\pm(2Mu^3-u^2+c^2)^{-1/2} \;,
\end{equation}
where $+$ corresponds to $d\varphi/du>0$, $-$ corresponds to
$d\varphi/du<0$, and $c=E/L$ for $L\neq0$. The case of $L=0$, which
corresponds to radial null geodesics, has been discussed in
Subsection \ref{radial}. It turns out that, for $L\neq0$, the
geodesic depends on $E$ and $L$ only through their ratio $c$. Using
Eq.\ (\ref{Transform}) and Eq.\ (\ref{Conserve}), we can further obtain
\be 
\frac{dt}{d\varphi}=\frac{cr^2}{1-2M/r}+\frac{dr/d\varphi}{r/2M-1} \;,
\ee 
which can be simplified by using the new variable $u=1/r$ and
Eq.\ (\ref{Uphi}) as
\begin{equation}\label{Tphi}
\frac{dt}{d\varphi}=\frac{c\mp 2Mu\sqrt{2Mu^3-u^2+c^2}}{u^2-2Mu^3} \;.
\end{equation}
Alternatively, we can put it into an equation involving only $t$ and
$u$,
\begin{equation}\label{Tu}
\frac{dt}{du}=\frac{\pm c(2Mu^3-u^2+c^2)^{-1/2}- 2Mu}{u^2-2Mu^3} \;,
\end{equation}
where, as mentioned before, $+$ corresponds to $d\varphi/du>0$, $-$
corresponds to $d\varphi/du<0$. Eq.\ (\ref{Uphi}) and Eq.\ (\ref{Tphi}),
or equivalently Eq.\ (\ref{Uphi}) and Eq.\ (\ref{Tu}) is a full set of
equations for generic null geodesics with non-zero angular momenta.

Now given $E_1$ and $E_2$ 
which do not satisfy
any of the sufficient conditions in Subsection \ref{sufficient
condition}, we have to do the following numerical calculation to see
if they are causally related to each other. For any $c$ we can
integrate Eq.\ (\ref{Uphi}) from $\varphi_1=0$, $u_1=1/r_1$ to
$u_2=1/r_2$, and get some value $\varphi_2'$. By choosing a suitable
$c_0$, we can make $\varphi_2' = \varphi_2$
which means that
the null geodesic with
$c_0$ hits $\gamma$ from $E_1$.  Then we can use $c_0$ in
Eq.\ (\ref{Tu}), and integrate it from $u_1$, $t_1$ to $u_2$ and get some
value $t$. If $t\leq t_2$, then they are definitely causally
related.  If $t > t_2$, according to the lemma of Section
\ref{lemma_section}, they must be causally unrelated.

\section{Results} 
\label{results}

\subsection{Causal Relations}
\label{table}

We sprinkle into a region of Schwarzschild spacetime, which is bounded by $0
\leq r \leq \rmax$ and $0 \leq t \leq \tmax$.  See Appendix
\ref{sprinkling-sec} for details on how this is done.
It is important to note that in this section we use the unrotated
coordinates, so $\theta$ is not restricted to $\frac{\pi}{2}$,
and use coordinates such that $M = 1$.  By `equatorial plane', we simply mean
that $\theta = \frac{\pi}{2}$.

Nine events selected from a region of Schwarzschild with $\rmax = 3,
\tmax = 8$, are shown in Table \ref{elts}.\footnote{We chose $\rmax = 3$ to get
  roughly the same number of interior and exterior events, and $\tmax = 8$ to
  be roughly half the circumference of a circle at $r=3$.  Note that $r=3$
  corresponds to the innermost circular orbit (which is therefore lightlike).}
\begin{table}[htbp]
\center
\begin{tabular}{|c|cccc|c|}
\hline
event & $t$ & $r$ & $\theta$ & $\phi$ & \\
\hline
0 & 0.410895 & 2.36161 & 1.80295 & 0.57951 & exterior\\
1 & 1.109415 & 2.89891 & 1.04335 & 4.25531 & exterior\\
2 & 1.133105 & 1.36083 & 1.89919 & 1.06482 & interior\\
3 & 2.743428 & 2.74093 & 2.97906 & 4.22204 & exterior\\ 
4 & 3.235970 & 0.65462 & 0.11664 & 5.06884 & interior\\
5 & 3.972871 & 0.96354 & 2.33727 & 1.38169 & interior\\ 
6 & 5.230757 & 2.34476 & 1.11855 & 3.47242  & exterior\\
7 & 6.014261 & 0.664739 & 2.82235 & 0.95459 & interior\\ 
8 & 6.193089 & 0.429636 & 2.20122 & 1.99644 & interior\\ 
\hline
\end{tabular}
\caption{Nine events in Schwarzschild spacetime, specified by their
  Eddington-Finkelstein coordinates.}
\label{elts}
\end{table}

Our task is to decide, for each pair of events $(E_1, E_2)$ (with $E_1$
having an earlier EF time coordinate than $E_2$), 
whether they are causally related or not.  To do this we perform the following
algorithm: 
\begin{enumerate}
\item Is $E_1$ is behind the horizon and $r_2 > r_1$?  If so
  they are unrelated.
\item Change the angular coordinates so that both lie on the equatorial plane,
and restrict attention to null geodesics which traverse an azimuthal angle
$\leq \pi$ on their trip from $E_1$ to a stationary worldline $\gamma$
containing $E_2$.
\item Is the EF time separation of the events less than the angular
  or radial spacelike bounds?  If so they are unrelated.
\item Is the EF time separation greater than the timelike bound?  If so they
  are related.
\item If neither sufficient condition is satisfied, then we must numerically
  compute the value of $(E/L)^2$ which will send a null geodesic from $E_1$ to
  $\gamma$.  Armed with this value, we compute the elapsed
  coordinate time along this geodesic, and 
  decide if it arrives before or after the event $E_2$.
\end{enumerate}

Table \ref{pairs} gives details for a selection of pairs of 
events from Table \ref{elts}.
In particular we show the various quantities which are
computed along the way to deciding if this pair is causally related.
\begin{table}[htbp]
\setlength{\tabcolsep}{1mm}
\begin{tabular}{|c|ccccccccc|l|}
\hline
pair & dir & $r_0$ & $\varphi_2$ & $\Delta t$ & rad trip & ang bnd & tot trip & $(E/L)^2$ & time & result\\
\hline
0 1 & out & 2.898906 & 2.567258 & 0.698520 & 4.179694 & 13.36484 & ---      & ---       & ---     & unrelated : either bound\\
0 2 & in  & ---      & 0.475629 & 0.722210 & 1.000779 & 0.647253 & ---      & ---       & ---     & unrelated : radial bound\\
1 2 & in  & ---      & 2.937662 & 0.023690 & 1.538071 & 3.997673 & ---      & ---       & ---     & unrelated : either bound\\
0 4 & in  & 2.361614 & 1.827161 & 2.825075 & 1.706999 & 1.196088 & 12.73426 & 0.0460462 & 4.69799 & generic, crossing, unrelated\\
1 4 & in  & 2.898906 & 0.965528 & 2.126556 & 2.244290 & 0.632050 & ---      & ---       & ---     & unrelated : radial bound\\
2 4 & in  & 1.360835 & 1.973603 & 2.102865 & 0.706219 & 1.291951 & ---      & ---       & ---     & generic, hits singularity\\
0 5 & in  & 2.361614 & 0.867289 & 3.561976 & 1.398076 & 0.835666 & 6.632326 & 0.272754  & 2.11915 & generic, crossing, related\\
1 5 & in  & 2.898906 & 2.821962 & 2.863456 & 1.935368 & 2.719068 & 16.62616 & 0.0388018 & 9.86353 & generic, crossing, unrelated\\
2 5 & in  & 1.360835 & 0.512295 & 2.839766 & 0.397297 & 0.493616 & ---      & ---       & ---     & generic, hits singularity\\
4 5 & out & \multicolumn{9}{r|}{emerging from interior : unrelated}\\
0 6 & in  & 2.361614 & 2.820685 & 4.819862 & 0.016859 & 6.613817 & ---      & ---       & ---     & unrelated : angular bound\\
1 6 & in  & 2.898906 & 0.690536 & 4.121342 & 0.554150 & 1.619139 & 4.148999 & 0.0476468 & 2.60973 & generic, exterior, related\\
2 6 & out & \multicolumn{9}{r|}{emerging from interior : unrelated}\\
4 6 & out & \multicolumn{9}{r|}{emerging from interior : unrelated}\\
5 6 & out & \multicolumn{9}{r|}{emerging from interior : unrelated}\\
3 7 & in  & 2.740935 & 0.480906 & 3.270833 & 2.076196 & 0.319677 & 4.611431 & 4.55373   & 2.1833  & generic, crossing, related\\
5 8 & in  & 0.963538 & 0.485959 & 2.220218 & 0.533902 & 0.208785 & ---      & 1.35646   & 0.667059& generic, interior, related\\
\hline
\end{tabular}
\caption{Considering relations between a collection of pairs of elements of
  Table \ref{elts}.}
\label{pairs}
\end{table}
For each pair, if applicable, we show:
\begin{itemize}
\item the direction (ingoing or outgoing)
\item the angle between the events $\varphi_2$
\item the time coordinate separation between the events $\Delta t$
\item the `radial' spacelike bounds, given by \eq{rad_bnd_ingoing} for
  ingoing null geodesics, and \eq{rad_bnd_outgoing} for outgoing geodesics
\item the `angular' spacelike bound, \eq{ang_bnd}
\item the coordinate time traversed along the trip, composed of successive
  segments at constant $r$ or angular position, which yields the
  timelike bound of Subsection \ref{timelike_bnd}
\item the value of $c^2 = (E/L)^2$ in \eq{Uphi} which yields the fastest null
  geodesic from $E_1$ to $\gamma$.  These values are computed using the
  procedure detailed in Appendix \ref{numint_details-sec}.
\item the coordinate time elapsed along this fastest geodesic
\item whether they are related or not, and which condition allows us to
  decide
\end{itemize}
For the 0 1 pair, the time coordinate separation $\Delta t$ is less than
either of the spacelike bounds, so the events are unrelated.  The 0 2 pair
fails the angular spacelike bound, but passes the radial bound.
The 0 4 pair
is `generic', meaning that it fails both spacelike bounds and the timelike
bounds.  These simple tricks are insufficient to determine if the events are
related, so we must integrate \eq{Uphi} to locate the fastest null geodesic
from event 0 to the $\gamma$ containing event 4.  This geodesic crosses the
horizon, but does not arrive at $\gamma$ in time for the events to be
related (since the `time' in the last column is larger than the `available
time' $\Delta t$).
The 2 4 pair fails the radial spacelike bound.  The radial and
timelike bounds fail because the pair is inside the horizon, for which there
is no timelike, constant-$r$ trajectory.  Integrating (\ref{Uphi}), with
$c^2\equiv (E/L)^2=0$, gives only $\varphi = 0.721811$, which is not enough
to reach $\gamma$.  All null geodesics will hit the singularity before
reaching the $\theta_2 = 0.11664, \phi_2 = 0.11664$ worldline.  
The 2 5 pair
meets a similar fate: there are no future directed null geodesics from event
2 which reach $\theta_2 = 2.33727, \phi_2 = 1.38169$ before falling into the
singularity.
The 0 5 pair fails all bounds, and thus is generic.  This time, however, the
null geodesic does reach $\gamma$ before event 5.
The 4 5 pair represents an attempt to `escape from the interior', in the
sense that event 4 is inside the horizon, and event 5 is at a larger radius
than 4.  Even though event 5 is also inside the horizon, there are no causal
curves inside the horizon which extend to larger radii.
Events 0 and 6 are at almost the same radius, but at very different angular
positions.  Thus the angular bound is useful in deducing that they are
unrelated, without having to integrate any geodesics.
Events 1 and 6 are an example of a generic pair which are both outside of the
horizon.  They happen to be related. 
The pairs 2 6, 4 6, and 5 6 suffer the same fate as 4 5.  This time they
are even attempting to escape across the horizon.
The 5 8 pair is generic and completely inside the horizon.  This time there
are causal curves which reach $\gamma$ from event 5, and the events end up being related.

\subsection{Causal Sets}

\begin{figure}[htbp]
\includegraphics[width=\textwidth]{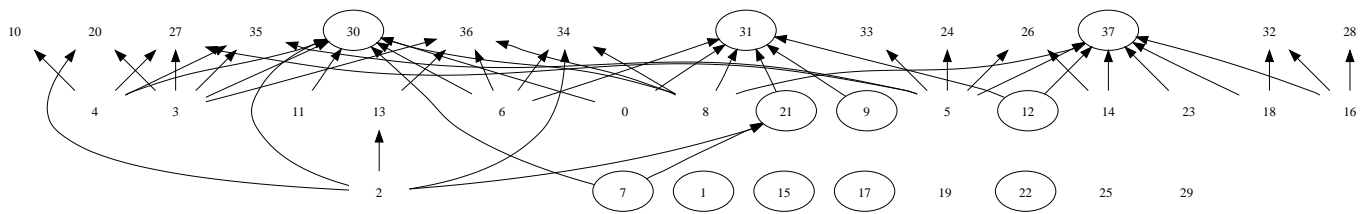}
\caption{Hasse diagram of a 38 element causal set sprinkled into
  Schwarzschild spacetime, with $\rmax=4, \tmax=8$.  The circled elements
  landed behind the horizon.  Note that no information escapes from this
  interior region, in that there are no causal relations from circled to
  uncircled elements.}
\label{hasse}
\end{figure}

\begin{figure}[htbp]
\includegraphics[width=\textwidth]{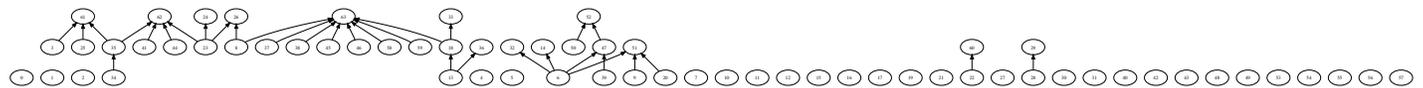}
\caption{A `causal set singularity'.  This 64 element causet is sprinkled
  near the singularity of Schwarzschild,
  $\rmax = 1, \tmax \to \infty$.  (The resulting causet is independent of
  $\tmax$ beyond some finite value.)}
\label{singularity}
\end{figure}

\begin{figure}[htbp]
\center
\includegraphics[width=\textwidth]{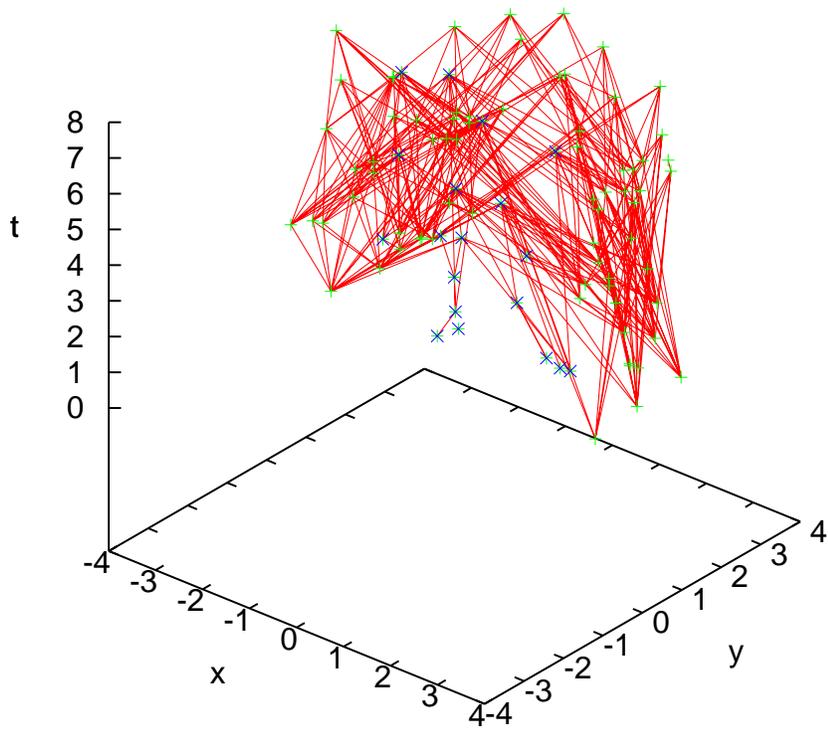}
\caption{Hasse diagram of a causal set $C$ which arises from sprinkling 91
  elements into the half equatorial plane.  Elements sprinkled behind the
  horizon appear blue, while those outside are green.} 
\label{half128}
\end{figure}

\begin{figure} [htbp]
\center
\includegraphics[width=\textwidth]{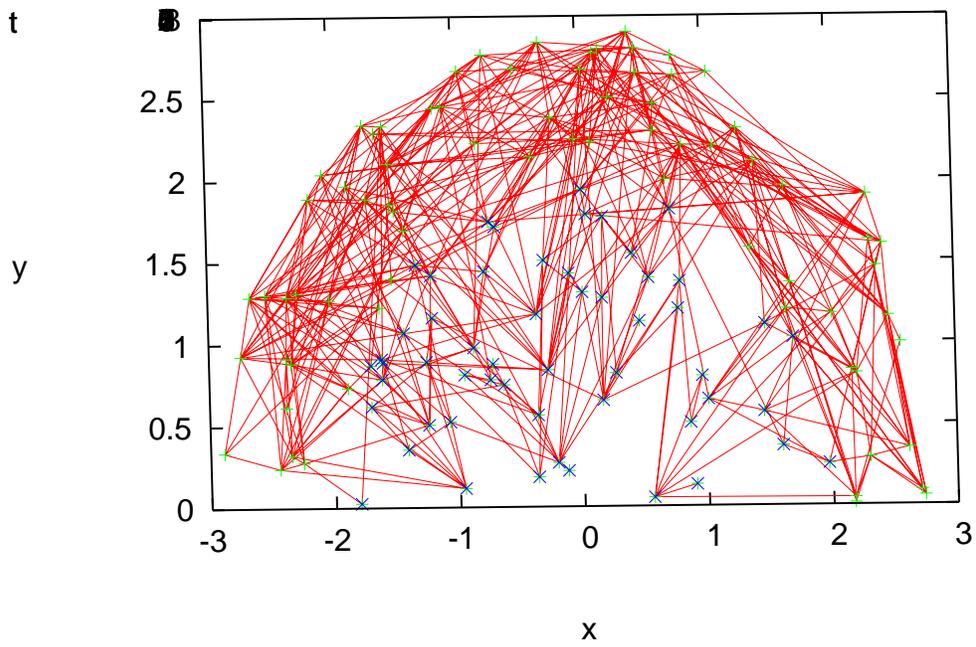}
\caption{The same 91 element causal set $C$
as viewed from the `top' ($t\to +\infty$, look into the past direction).  The
blue `half disk' corresponds to the black hole interior.}
\label{half128-top}
\end{figure}

\begin{figure}[htbp]
\center
\psfrag{phi}{\large $\phi$}
\includegraphics[width=\textwidth]{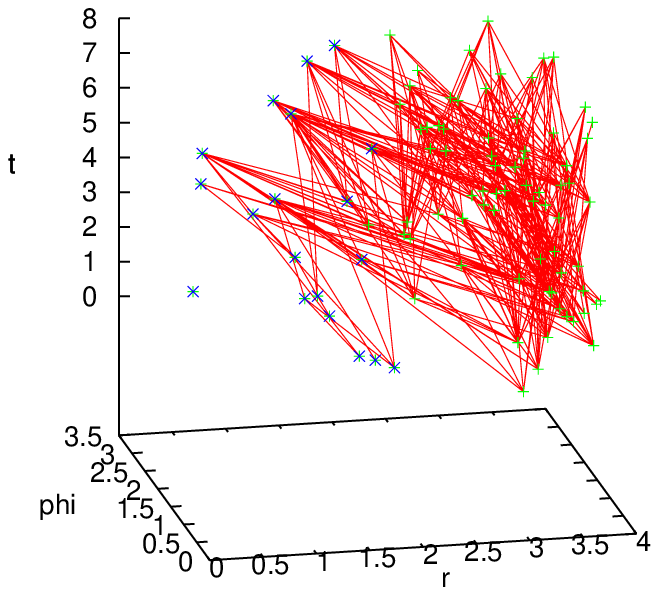}
\caption{The sprinkled 91 element causal set $C$, this time plotting $r$
  against $\phi$ interpreted as Cartesian coordinates, rather than polar
  coordinates.  The blue region on the left corresponds to the interior of
  the black hole.}
\label{half362-r-phi}
\end{figure}

\begin{figure}[htbp]
\center
\psfrag{phi}{t}
\psfrag{r2}{$r^2$}
\includegraphics[width=\textwidth]{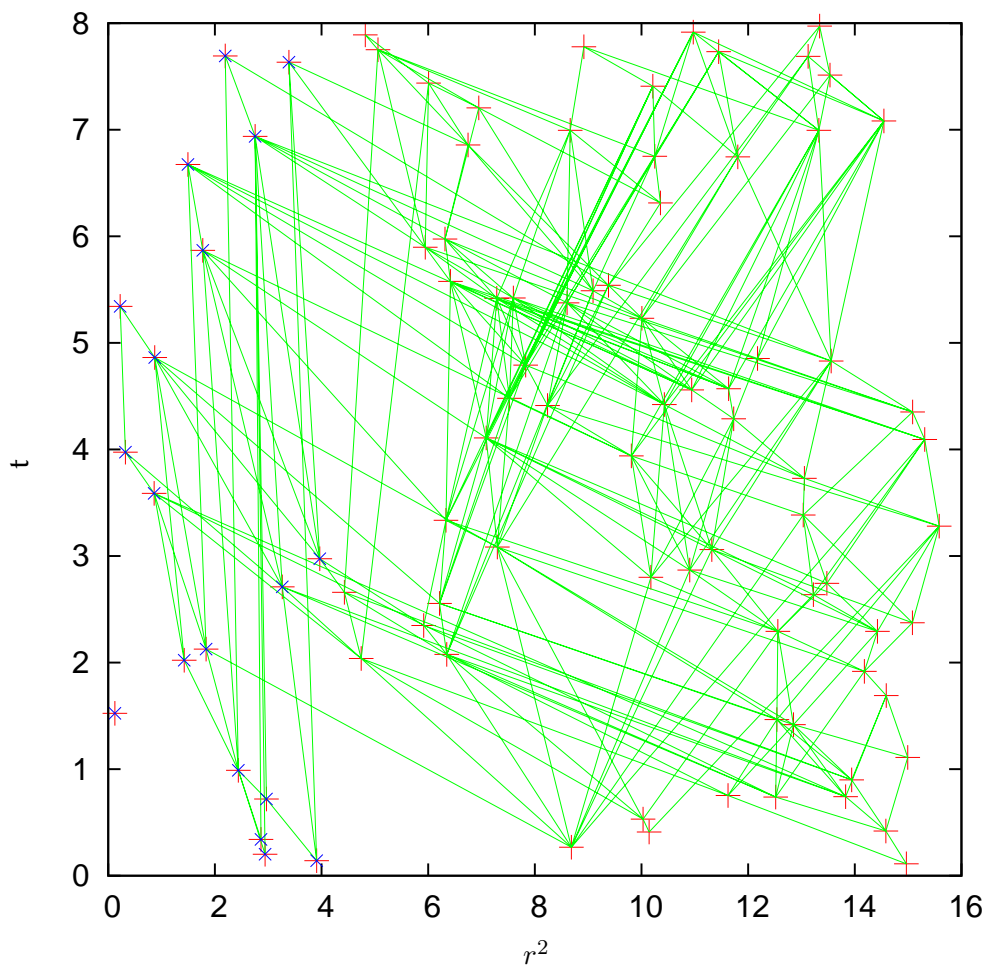}
\caption{Sprinkling of 91 elements into a 1/10 radian wedge of the
  equatorial plane.  Here we ignore the $\phi$ coordinate, and square the
  radial coordinate in order to spread out the elements at large radius.}
\label{wedge}
\end{figure}

In this subsection we show some Hasse diagrams of the causal sets which arise
from sprinkling into Schwarzschild.  In the Hasse diagram one shows only the
\emph{links} of the causal set, namely those causal relations which are not
implied by transitivity.  Figures \ref{hasse} and \ref{singularity} use the
graphviz package \cite{graphviz} to generate the diagram, ignoring the
embedding information.  Figures \ref{half128}, \ref{half128-top},
\ref{half362-r-phi}, and \ref{wedge} arrange the causal set elements using
their embedded location.

Figure \ref{hasse} shows a 38 element causal set, which arises from a
sprinkling into a region of a Schwarzschild spacetime.  Figure
\ref{singularity} portrays a causal set which arises when sprinkling near the
singularity.  Note that there is no need to worry about sprinkling on top of
the singularity, as that is a zero probability event, even if the singularity
is contained within the sprinkling region.
It is very antichain-like, as the futures of the elements rapidly fall into
the singularity, so it is unlikely that another sprinkled element lands in
that region.

In the remaining four figures of this section we show more Hasse diagrams, though this time
we use the embedding information to locate the nodes in the graph.  For the
first three, Figures \ref{half128}, \ref{half128-top}, and
\ref{half362-r-phi} we generate a causal set $C$ by sprinkling into half of
the equatorial plane ($\theta = \pi/2$, $\phi \in [0, \pi]$), with $\tmax=8$
and $\rmax=4$.  The relations (links) are shown in red.  The half plane is
chosen
to reduce the `clutter' from links between distant elements (such as links
which cross the black hole, from the exterior region on one side to the
exterior region on the other).  Note
that these red lines are not null geodesics connecting the elements (though
they will converge to such in the infinite sprinkling density limit).  They
are simply straight lines between the elements drawn by the plotting program
(gnuplot).

The last figure \ref{wedge} attempts to more clearly show the light cone
structure of Schwarzschild, in particular the tilting of the light cones
toward the singularity at $r=0$.  Here we sprinkle 91 elements into a 1/10
radian `wedge' of the equatorial plane.
Since the number of elements sprinkled into a region grows quadratically with
$r$, we plot here $t$ versus $r^2$, so the distribution will appear
approximately uniform in the horizontal direction.  On the right side of the
figure we have a cutoff at $r=4$, so obviously no links can go beyond that
cutoff.  However around $r=3$ we can see the light cone spreading in both
directions.  As we get closer to the center the links extend less and less to
the right (exterior).  As we cross the horizon at $r=2$ we see that the links
only extend to the left, since their are no future directed causal curves
which have non-decreasing $r$ in the interior.

\section{Conclusion and discussion}
\label{conclusion}

We have described an algorithm to determine if two events are causally
related in Schwarzschild spacetime.  It involves first checking a number of
sufficient conditions, to see if these are able to determine whether they are
related, without resorting to a time consuming numerical integration.  If
none of these are satisfied then we locate a future directed null geodesic
which leads from the earlier event to a worldline containing the later, and integrate the
elapsed EF time along the geodesic to determine if the pair is related.

We then wrote a `thorn' (module) within the Cactus framework with sprinkles
events into a region of Schwarzschild, and use this algorithm to deduce the
causal relations between every pair of events.  This procedure yields a
causal set which `faithfully embeds' into a Schwarzschild black hole spacetime.

It should not be difficult to generalize this prescription to other black
hole spacetimes, such as Reissner-Nordstrom or Kerr.  In order to describe
the closed timelike curves in interior regions one may generalize the
definition of the causal set slightly, replacing the irreflexive condition
with reflexivity, so one ends up with a transitive directed graph. 
In addition one can consider other conformally
curved spacetimes, such as the $k=\pm 1$ Friedman-Robertson-Walker
universes.\\

This work is a good starting point for one to address a wide
array of questions within the causal set program, which previously were
inaccessible.  On one hand it will allow one to easily investigate many
kinematical questions with regard to the so-called `Hauptvermutung' of causal
sets, which conjectures that if one has a causal set which is likely to arise
from sprinkling into two separate spacetimes, then those spacetimes must be
approximately isometric.  To date, all results regarding to how to deduce
properties of an approximating continuum from a causal set consider only
conformally flat geometries.  Now one will be able to test such constructs on
a much wider class of geometries.

Another important application is towards our understanding of black hole
thermodynamics.  The fundamental discreteness of causal sets gives us a
possible way to characterize the degrees of freedom which give rise to black
hole entropy.  
We are now in a position to repeat the analysis of the link
counting and its generalizations \cite{bhentropy}, for the full 4d
Schwarzschild geometry.  As argued by Dou and Sorkin, this may provide access
to the fundamental length scale of quantum gravity, since entropy, as a pure
number, is not subject to renormalization.

Finally, a long standing question in semi-classical gravity is the 
trans-Planckian problem, that the Hawking radiation emanating from a black
hole horizon, at late times (after the black hole forms), arises from modes
of frequency much greater than the Planck scale \cite{hawking_radiation}.
If spacetime is discrete at
this scale, then one may expect that such trans-Planckian modes cannot exist.  How
then is Hawking radiation possible, in such a discrete setting?
Important groundwork on the dynamics of scalar fields on a background causal
set has recently been laid \cite{dalembertian, johnston, sverdlov}.
Now that we can construct causal sets which correspond to a full four dimensional black
hole, it may be possible to address this question within the causal set
approach.

\subsection*{Acknowledgments}
We are extremely grateful to Joseph Samuel and Rafael Sorkin for illuminating
discussions on the behavior of null geodesics and their relation to causal
structure.  SH also thanks Hongbao Zhang for many
helpful discussions and comments.

This research was supported by the Perimeter Institute for Theoretical Physics.
Research at Perimeter Institute is supported by the Government of Canada
through Industry Canada and by the Province of Ontario through the Ministry of
Research \& Innovation.  SH was supported by NFSC grants 10235040 and 10421003.

SH thanks the Perimeter Institute for hospitality while this work was carried out.

\appendix
\section{Appendix: Proof of the proposition}
\label{proof}
In this section, we want to prove the following proposition. Given
two events $E_1$ and $E_2$ in Schwarzschild spacetime, with
$t_1<t_2$, and $E_2$ representing the moment $t_2$ on the world line
$\gamma$ of a stationary observer, the null geodesic with
$|\Delta\varphi|=\varphi_2$ is the fastest one (arrives at the earliest time in $EF$ coordinates)
for all future-directed null geodesics from $E_1$ to $\gamma$.

It is clear that the fastest geodesic will have the least elapsed time.  Thus
from \eq{Tu}, we wish to minimize the integral
\begin{equation}
\Delta t = \int_{u_1}^{u_2} \frac{\pm c(2Mu^3-u^2+c^2)^{-1/2}- 2Mu}{u^2-2Mu^3} du
\label{dt}
\end{equation}
along the geodesic.  Since $d\varphi$ is clearly nonnegative, the sign in
front of $c$ will be positive for ingoing geodesics ($du>0$) and negative for
outgoing geodesics ($du < 0$).
Given $E_1$ and $E_2$ (and thus $\gamma$), different null geodesics are given by different values
of $c$, and it is straightforward to see from Eq.\ (\ref{dt}) that in every
case for larger $c$, $\Delta t$ gets smaller. Therefore, the fastest null
geodesic from $E_1$ to $\gamma$ must have the largest possible $c$ to reach
$\gamma$, which, by Eq.\ (\ \ref{Uphi}), has the smallest possible $|\Delta
\varphi|=\varphi_2$.\footnote{As mentioned earlier, for
  $\varphi_2<\pi$, there is only one fastest null geodesic with the largest
  $c$ to make $\Delta\varphi=\varphi_2$; while for $\varphi_2=\pi$, there are
  in fact two fastest null geodesics with $\Delta\varphi=\pm\pi$ but the same
  $c$ and $\Delta t$.}

\section{Numerical Details}
\label{details}

\subsection{Sprinkling into Schwarzschild Spacetime}
\label{sprinkling-sec}
In implementing these ideas on a computer, the first task is to randomly
select events in spacetime, with a density proportional to the spacetime
volume factor
\bne
\sqrt{-g} = r \sqrt{r^2 + 12 M^2} \sin \theta \;.
\label{vol_elt}
\ene
Because of the simple product form of this expression, we can break up the
sprinkling into an angular piece, a temporal piece, and a radial piece.
For the uniform sprinkling of the angular 
coordinates, on a 2-sphere, we follow the procedure described in Section
5.2 of \cite{NumVolSpec}.  For the temporal sprinkling,
since the volume element is independent of $t$, we can select its values
uniformly at
random.

The sprinkling in the radial direction must be performed such that it yields
the distribution of \eq{vol_elt} (ignoring the $\theta$ dependence; this is
accounted for above).  This is achieved by the following general method 
(derived from \cite{knuth}).  To sprinkle a coordinate $x$ between the bounds
$a \leq x \leq b$ such that it has
distribution $f(x)$, compute the indefinite integral
\bne
I(x) = \frac{1}{N} \int_a^x f(x') dx' \;,
\label{Iofx}
\ene
where the constant normalization factor is
\be
N = \int_a^b f(x') dx' \;.
\ee
Now invert \eq{Iofx} to get $x$ as a function of $I$.
This expression, 
where $I$ is a random variable distributed
uniformly in the unit interval $[0,1]$, will be distributed according to
$f(x)$.  Of course this method only works if $f(x)$ is integrable and its
integral is invertible.  In our case $f(r) = r \sqrt{r^2 + 12 M^2}$, and the
expression is
$\sqrt{(3NI + (a^2+12M^2)^{3/2})^{2/3} - 12M^2}$.
\footnote{In fact the angular sprinkling is of this type as well.  With
  $f(\theta) = \sin \theta$ the expression is $\arccos(2I-1)$.  The $\phi$
  coordinate is distributed uniformly in $[0,2\pi]$.}

\subsection{Determining the Causal Relations}
\label{numint_details-sec}
Once equipped with a collection of events in Schwarzschild, we can sort them
by their time coordinate, and then consider each sorted pair in turn.  The
first task is to check the sufficient conditions described in Subsection
\ref{sufficient condition}.  This is relatively straightforward; numerous
examples are given in
Subsection \ref{table}.  If the pair fails all
available sufficient conditions, then it is a `generic pair', and we must
integrate null geodesics as described in Subsection \ref{generic pairs}.

\begin{figure}[htbp]
\center
\psfrag{u}{\cmt{1}{\vspace*{.4mm} $Mu$}}
\psfrag{f(u)}{$M^2 f(u)$}
\includegraphics[width=13cm]{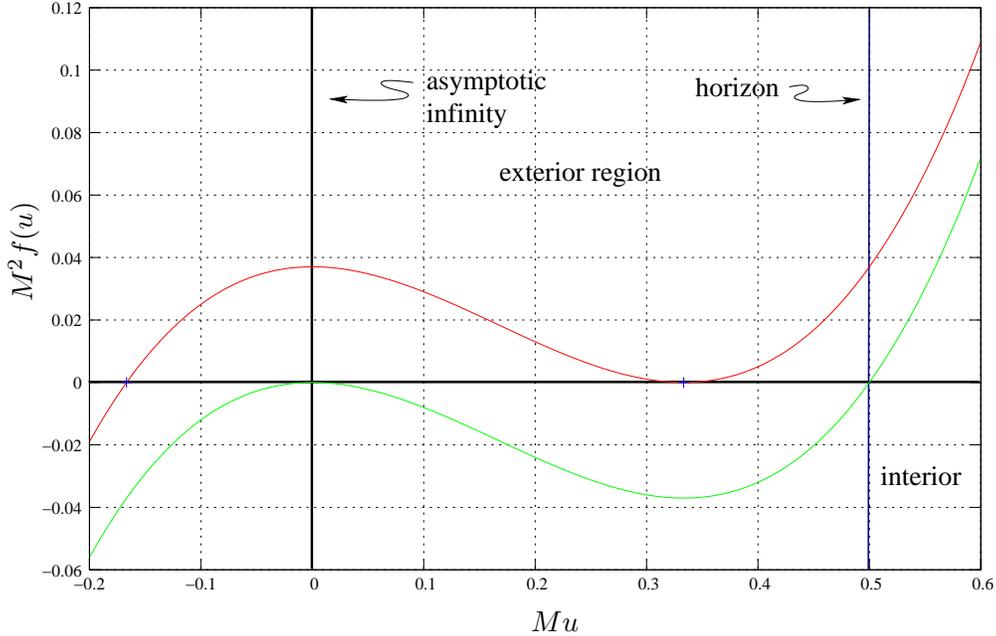}
\caption{Cubic $f(u)=2Mu^3-u^2+c^2$ in denominator of \eq{Uphi}.  The red
  curve has $c^2 = \frac{1}{27M^2}$, while the green has $c^2=0$.}
\label{cubic}
\end{figure}

The basic task is to find a value of the parameter $c^2=(E/L)^2$ for which
the integral of \eq{Uphi} along the null geodesic from $u_1$ to $u_2$ equals
$\varphi_2$.  This is made complicated by the fact that the cubic
$f(u)=2Mu^3-u^2+c^2$ in the denominator of the right hand side can have real
roots within the domain of integration.  (See Fig.\ \ref{cubic} for an
illustration.)
The behavior of these roots is as follows.  For $c^2 > \frac{1}{27M^2}$, the
cubic has no non-negative real roots.  At $c^2 = \frac{1}{27M^2}$ it has a
double root at $u=\frac{1}{3M}$ (and a third at $u=\frac{-1}{6M}$, which is
irrelevant because we consider only non-negative values of $u=1/r$).  This
double root, which causes the integrand to diverge, corresponds to the pure
angular orbit at $u=\frac{1}{3M}$.  As we further reduce $c^2$,
the double root at
$u=\frac{1}{3M}$ separates into two, the larger increasing toward
$u=\frac{1}{2M}$ (the horizon), and the smaller decreasing toward $u=0$
(asymptotic infinity), which they reach when $c^2$ descends to its smallest
possible value 0.
So the task of our algorithm is to shoot null geodesics from $E_1$ in
different directions (different values of $c^2$), trying to hit $\gamma$
($\varphi_2$), all the time being careful to keep $c^2$ large enough that the
roots of the cubic do not fall between either $u_1$ or $u_2$, as doing so
would cause the integrand to become infinite or imaginary.

Our algorithm begins by setting $c^2=0$ if either event lines behind the
horizon, or $c^2 = \frac{1}{27M^2}$ otherwise.  It will later adjust the
value of $c^2$, in its attempt to locate a null geodesic which travels from
$E_1$ to $\gamma$.  For each value of $c^2$, it first checks to see if either
$u_1$ or $u_2$ lies between the non-negative real roots (if any).  If so, it
adjusts $c^2$ upward (by using a linear extrapolation of $f(u)$ at the
roots),
until neither $u_1$ or $u_2$ lies between them.

Now that we have an integrand which is real and finite in the entire domain
$[u_1,u_2]$, the code integrates \eq{Uphi} numerically using the composite
Simpson's rule, with $n=512$ subintervals. 
If either the
numerically evaluated integral
is within $\Delta=5\epsilon$ of $\varphi_2$, where $\epsilon$ is double
precision machine epsilon ($\approx 2.22045 \times 10^{-16}$ on the machine
on which the results of section \ref{results} were generated),
or the two most recently chosen values of $c^2$ are within $5\epsilon$ of
each other (see below),
we then (tentatively) decide that this null geodesic arrives at
$\gamma$.  If this condition fails,
we choose a second value for $c^2$ and repeat.
The second value is chosen to be the the first + .03 if the integral
overshoots (is greater than) $\varphi_2$, or the first - .005 if it
undershoots.  
If the second guess of $c^2$ also misses $\varphi_2$, then subsequent
values are chosen by a linear interpolation/extrapolation from the two
previous guesses.  This algorithm (with the additional features described
below) converges for all pairs of events we have
encountered in our simulations.

There is a special situation which can arise (as in some of the examples of
Subsection \ref{table}), in which there are \emph{no} causal curves from
$E_1$ to $\gamma$.  This occurs when
$E_2$ lies behind the horizon, and the integral with $c^2=0$ undershoots
$\varphi_2$. This means that every future directed causal curve from $E_1$
falls into the singularity before reaching $\gamma$, so the events must be
unrelated.

When $c^2$ is small enough that the domain of integration touches a root, the
integral diverges.  Often the `target value' of $\varphi_2$ requires a $c^2$
which is is very close to this singular value.  We find that a convenient way
to handle this situation numerically is to detect when we manage to find a
valid value of $c^2$, which is large enough for $u_1$ and $u_2$ to escape the
roots, and yet small enough to yield an integral which exceeds $\varphi_2$.
Once we find this value of $c^2=c_{\mathrm{min}}^2$, then we know that the
value of $c^2$ we seek is greater than this.  Thus, in the course of the
above iteration, which uses linear interpolation/extrapolation to select
subsequent values of $c^2$, if a value is selected which is smaller than
$c^2=c_{\mathrm{min}}^2$, then we instead choose the mean of
$c^2=c_{\mathrm{min}}^2$ and the previous $c^2$.  (Furthermore in subsequent
iterations, if we find a yet larger value of $c^2$ for which the integral
exceeds $\varphi_2$, then we use this as the new $c_{\mathrm{min}}^2$.)

Once the above loop converges, so that we have a value of $c^2$ for which the
integral of \eq{Uphi} yields $\varphi_2$, we then check that the numerical
approximation to the integral is sufficiently accurate.  The check is simple:
we compute the numerical approximation to the integral again at four times
the resolution $\varphi_2(4n)$ ($4n$ subintervals), and subtract that value
from the $\varphi_2(n)$ using $n$ subintervals.  If the difference is greater
than $8 \eta$, where $\eta$ is the larger of $\varphi_2(4n) - \varphi_2$ and
$5\epsilon$, then we double $n$ and repeat the above iteration.  (Though we
stop the iteration if the difference $|\varphi_2(4n) - \varphi_2(n)|$
ever \emph{increases} from that for the previous $n$.)

Now that we have an accurate value of $c^2$, which yields a null geodesic
which hits $\gamma$, we integrate \eq{Tu} to get the elapsed time along the
geodesic, and thus can determine if the two events are related by comparing
this elapsed time with $t_2 - t_1$.

\end{document}